\documentclass{PoS}

\usepackage{braket,marvosym}

\renewcommand{\d}{\ensuremath{\mathrm{d}}}

\renewcommand{\d}{\ensuremath{\mathrm{d}}}

\newcommand{\p}{\partial}

\title{Ghost dissection}

\ShortTitle{Ghost dissection}

\author{\speaker{David Dudal} \& Nele Vandersickel\\
        Ghent University, Department of Physics and Astronomy, Krijgslaan 281-S9, 9000 Gent, Belgium\\
        E-mail: \email{david.dudal@ugent.be,nele.vandersickel@ugent.be}}

\author{Attilio Cucchieri \& Tereza Mendes\\
        Instituto de F\'\i sica de S\~ao Carlos, Universidade de S\~ao Paulo, Caixa Postal 369, 13560-970 S\~ao Carlos, SP, Brazil\\
        E-mail: \email{attilio@ifsc.usp.br,mendes@ifsc.usp.br}}

\abstract{We show that a necessary condition to have a positive Landau-gauge ghost propagator in $d=2$ Yang-Mills theories is a vanishing zero-momentum gluon propagator. Our proof is based on a careful scrutinizing of the ghost Dyson-Schwinger equation. Said otherwise, the Gribov no-pole condition forbids the occurrence of the ``decoupling/massive'' gluon propagator solution in $d=2$, in sharp contrast with $d=3$ and $4$, but consistent with state-of-the-art lattice data.  }

\FullConference{International Workshop on QCD Green's Functions, Confinement and Phenomenology,\\
		September 05-09, 2011\\
		Trento Italy}

\begin{document}

\section{The Gribov no-pole condition}
In the seminal work \cite{LENINGRAD-77-367} Gribov noticed that neither the (covariant) Landau nor the (noncovariant) Coulomb gauge are suitable to select a single gauge-field representant on each orbit of gauge-equivalent fields. Later on, it was recognized that this so-called Gribov ambiguity is a generic problem for non-Abelian gauge fixing \cite{135165}.

The main observation by Gribov is that, upon considering an infinitesimal gauge transformation parameterized via $A_\mu^{a \prime }=A_\mu^a +D_\mu^{ab}\omega^b$, one observes that the Landau-gauge condition, $\p_\mu A_\mu=0$, is also fulfilled by the equivalent field, i.e.\ $A_\mu^\prime$, $\p_\mu A_\mu^\prime=0$, if the Landau-gauge Faddeev-Popov operator
\begin{equation}
\mathcal{M}^{ab}(x,y) \, = \, - \delta(x-y) \, \p_\mu  D_\mu^{ab}
                      \, = \, \delta(x-y) \, \left( - \p_\mu^2 \delta^{ab} +  f_{abc}\p_\mu  A_\mu^c \right)
\end{equation}
has zero modes, since we must have, at first order in $\omega$,
\begin{equation}\label{mab2}
0= \p_\mu A_\mu^\prime = \p_\mu A_\mu - \mathcal{M}\omega = - \mathcal{M}\omega\,.
\end{equation}
If one removes the possibility of such zero modes by modifying the path integral measure, at least infinitesimally-related gauge copies will be ignored in any computation of expectation values. Let us also notice here that the existence of zero modes of the Faddeev-Popov operator invalidates, strictly speaking, the mathematical correctness of the textbook Faddeev-Popov quantization. Indeed, the latter procedure is based on the appropriate functional generalization of the identity
\begin{equation}
\left\{\left|\frac{\p f}{\p x}\right|\right\}_{x=x^\ast,f(x^\ast)=y}     \int \d x\; \delta[f(x)-y]\;g(x)= \left\{g(x)\right\}_{x=x^\ast,f(x^\ast)=y} \; ,
\end{equation}
where $\left.\frac{\p f}{\p x}\right|_{x=x^\ast}$ plays the role of the Faddeev-Popov determinant. However, the above expression is only valid if $f(x)=y$ has a \emph{single} solution at $x=x^\ast$, thereby also assuming that $\left.\frac{\p f}{\p x}\right|_{x=x^\ast}\neq0$. Additionally, assuming a positive Faddeev-Popov determinant also allows to drop the absolute value operation, something which is usually done. If $f(x)=0$ would have multiple solutions, let's say at $x=x_i^\ast$, one should use
\begin{equation}
 \int \d x\, \delta[f(x)-y]g(x)=\sum_i \left\{g(x)\left|\frac{\p f}{\p x}\right|^{-1}\right\}_{x=x_i^\ast}\,,
\end{equation}
but, to our knowledge, the functional generalization of this expression cannot be brought into a workable action for a gauge fixed SU(N) Yang-Mills theory, mainly because (1) a single Faddeev-Popov determinant can be lifted up into the action by the introduction of ghost fields, but a sum of determinants does not allow for this trick, and (2) the sign of the determinant should be taken into due account.

So, if one desires to stay as close as possible to the conventional Faddeev-Popov procedure, one better restricts the gauge-field region, over which the path integral runs, to the following set of configurations:
\begin{equation}\label{defgribovregion}
\Omega \, \equiv \, \left\{ \, A^a_{\mu}(x) : \, \p_{\mu} A^a_{\mu}(x)=0 \, , \, \mathcal{M}^{ab}(x,y) > 0 \right\} \; .
\end{equation}
This Gribov region, with the (first) Gribov horizon being its boundary $\p\Omega$, exists out of the relative minima under gauge transformations of the functional $\mathcal{E}[A]=\int \d^d x\, A_\mu^2$.
Gribov considered the ghost propagator written as
\begin{equation}
\mathcal{G}(k^2) \, = \, \frac{1}{k^2} \frac{1}{1- \sigma(k^2)} \;.
\end{equation}
By imposing
\begin{equation}
\sigma(k^2) < 1 \qquad \mbox{for} \qquad  k^2>0
\label{eq:nopole}
\end{equation}
he required \cite{LENINGRAD-77-367} that the ghost dressing function $k^2 \mathcal{G}(k^2)$ cannot have a pole
at finite nonzero momenta. Since the ghost propagator is actually given by
\begin{equation}
\mathcal{G}(k^2) = \frac{\delta^{ab}}{N^2 - 1} \, \Braket{ \, k \, \left| \left(\mathcal{M}^{-1}\right)^{ab} \right| \, k \, }  \; ,
\end{equation}
it is clear that the inequality (\ref{eq:nopole}), known as the no-pole
condition, should be equivalent to the restriction of the functional integration to the Gribov region $\Omega$.

In this proceeding, based on the work \cite{new}, we shall examine, as general as possible, the consequences of the no-pole condition; in particular, we shall focus on the specific case that the number of space-time dimensions is 2.  Why our interest in this case? At present day, there is little doubt that lattice simulations of the Landau gauge prefer the so-called decoupling/massive solution, in which case the gluon propagator attains a finite nonzero value at zero momentum, while the ghost propagator strives to a $1/p^2$ singularity at low momenta; this at least in 3 or 4 space-time dimensions \cite{Cucchieri:2007md,Sternbeck:2007ug,Cucchieri:2007rg,Bogolubsky:2009dc,Cucchieri:2008fc,Dudal:2010tf,Cucchieri:2011ig}. This was also discussed from the analytical viewpoint in e.g.~\cite{Aguilar:2004sw,Dudal:2008sp,Dudal:2011gd,Boucaud:2008ji,Aguilar:2008xm,Fischer:2008uz}. These results replaced the for a while rather popular scaling scenario, in which the gluon propagator vanishes at zero momentum, while the ghost propagator becomes more singular than $1/p^2$ \cite{Fischer:2008uz,Alkofer:2000wg}. The physical relevance of this propagator research program lies in their use in phenomenological applications, next to trying to gain a better understanding of confinement, or chiral symmetry breaking, in the Landau gauge. We refer to several other contributions in this proceedings to illustrate this.

In contrast with the $d=3,4$ cases, the scaling scenario keeps being supported by the $d=2$ lattice data, as is visible from e.g.~\cite{Cucchieri:2007md,Maas:2007uv,Cucchieri:2011um}. A careful analysis based on rigorously derived bounds allowed to show that $\mathcal{D}(0)=0$ persists in the infinite-volume limit \cite{Cucchieri:2007rg} (see also \cite{Cucchieri:2011um}).  Here, we present a summary of the detailed analysis presented in \cite{new}. The aim is to unravel a theoretical reason behind the impossibility to find a gluon propagator with $\mathcal{D}(0)>0$ in $d=2$. We shall see that the problem with a massive kind of gluon propagator always reveals itself under the form of a logarithmic infrared singularity in the ghost self-energy, viz.~the no-pole function $\sigma(k^2)$. Evidently, such singularity should be excluded, as it clearly violates the Gribov condition\footnote{In any case, we do not expect to find a singular term in the quantum correction $\sigma(k^2)$.} (\ref{eq:nopole}). For completeness, we also performed a similar analysis in $d=3,4$, in which case we are not forced to conclude that $\mathcal{D}(0)=0$, meaning that a decoupling/massive gluon is perfectly possible in $d=3,4$, and not necessarily in contradiction with the requirement (\ref{eq:nopole}).

\section{Scrutinizing the $d=2$ ghost propagator}
\subsection{Infrared singularity in and bounds on $\sigma(k^2)$ using one-loop perturbation theory }
As a warming up exercise, let us first consider the one-loop-corrected ghost propagator in the Landau gauge, which reads
\begin{equation}
\mathcal{G}(k^2) \, = \, \frac{1}{k^2} \, - \, \frac{\delta^{ab}}{N^2 - 1} \,
                          \frac{1}{k^4} \, g^2 f^{adc} f^{cdb} \int \frac{\d^d q}{(2\pi)^d}
                             \, (k - q)_{\mu} \, k_{\nu} \, \mathcal{D}(q^2) \, P_{\mu \nu}(q) \, \frac{1}{(k - q)^2} \; ,
\label{eq:Gini}
\end{equation}
with $ \delta^{ab} \, \mathcal{D}(q^2) \, \mathcal{P}_{\mu\nu}(q) $ the Landau-gauge gluon propagator and $\mathcal{P}_{\mu\nu}(q) = \left(\delta_{\mu\nu}-q_\mu q_\nu/q^2\right)$ the trans\-versal projector. For the moment, we also assume a tree-level ghost-gluon
vertex $i g f^{adc} k_{\nu}$. The expression (\ref{eq:Gini}) can be simplified and rewritten into the following form
\begin{equation}
\mathcal{G}(k^2) \, = \, \frac{1}{k^2} \frac{1}{1- \sigma(k^2)} \; ,
\label{eq:finalG}
\end{equation}
where $\sigma(k^2)$ is the momentum-dependent function given by
\begin{equation}\label{sigma1}
\sigma(k^2) \, = \, g^2 N \frac{k_{\mu} k_{\nu}}{k^2} \int \frac{\d^d q}{(2\pi)^d} \frac{1}{(k-q)^2} \mathcal{D}(q^2)
                        \, \mathcal{P}_{\mu\nu}(q) \; .
\end{equation}
This corresponds to the usual resummation of an infinite set of diagrams into the 1PI ghost self-energy. Notice that this resummation only makes sense when $\sigma(k^2)<1$, i.e.~when the no-pole condition (\ref{eq:nopole}) is satisfied.

We can then manipulate\footnote{We refer to \cite{new} for an alternative derivation.} $\sigma(k^2)$ by two consecutive partial integrations
\begin{eqnarray}
\frac{\sigma(k^2)}{g^2 N} & = & \frac{1}{8 \pi} \left[ \, \int_0^{k^2} \frac{\d x}{k^2} \, \mathcal{D}(x) \, + \,
                                       \int_{k^2}^{\infty} \frac{\d x}{x} \, \mathcal{D}(x) \, \right] \nonumber\\[2mm]
                          & = & \frac{1}{8 \pi} \left[ \, \frac{\hat{D}(k^2) - \hat{D}(0)}{k^2} \, + \,
                                       \ln\left( x \right) \, \mathcal{D}(x) \, \Bigr|_{k^2}^{\infty}  \, - \,
                                       \int_{k^2}^{\infty} \d x \, \ln\left( x \right) \, \mathcal{D}'(x) \, \right] \nonumber\\[2mm]
                          & = & \frac{1}{8 \pi} \left[ \, \frac{\hat{D}(k^2) - \hat{D}(0)}{k^2} \, - \,
                                       \ln\left( k^2 \right) \, \mathcal{D}(k^2) \, - \,
                                       \int_{k^2}^{\infty} \d x \, \ln\left( x \right) \, \mathcal{D}'(x) \, \right] \nonumber\\
&=&                                \frac{1}{8 \pi} \Biggl\{ \, \frac{\hat{D}(k^2) - \hat{D}(0)}{k^2} \, - \,
                                       \ln\left( k^2 \right) \, \mathcal{D}(k^2)  \, - \,
                  \left[ \, x \, \ln\left( x \right) \,-\,  x\, \right] \, \mathcal{D}'(x) \, \Bigr|_{k^2}^{\infty}
                                       \nonumber\\&& + \,  \int_{k^2}^{\infty} \d x \, \left[ \, x \, \ln\left( x \right) \,-\,  x\, \right]
                          \, \mathcal{D}''(x) \, \Biggr\} \nonumber\\[2mm]
                         & = & \frac{1}{8 \pi} \Biggl\{ \, \frac{\hat{D}(k^2) - \hat{D}(0)}{k^2} \, - \,
                                       \ln\left( k^2 \right) \, \mathcal{D}(k^2)  \, + \,
                  \left[ \, k^2 \, \ln\left( k^2 \right) \,-\,  k^2\, \right] \, \mathcal{D}'(k^2)\nonumber\\
                                        &&+ \,  \int_{k^2}^{\infty} \d x \, \left[ \, x \, \ln\left( x \right) \,-\,  x\, \right]
                          \, \mathcal{D}''(x) \, \Biggr\}\nonumber\,,
\label{form2}
\end{eqnarray}
where we assumed\footnote{More mathematical details, as well as a more profound discussion on the assumptions made at the level of the large/small momentum behaviour of the gluon propagator can be found in \cite{new}.} for simplicity that $\mathcal{D}(x) \sim 1/x$ at large $x$ and that also $\mathcal{D}'(x)$ goes to zero sufficiently fast at large momenta, e.g.~as $1/x^2$. We have denoted with $\hat{D}(k^2)$ a primitive of $\mathcal{D}(k^2)$. In the limit $k^2 \to 0$ we then obtain
\begin{equation}
\frac{\sigma(0)}{g^2 N} \, = \, \frac{1}{8 \pi}  \, \Biggl\{ \, \mathcal{D}(0) \, - \,
                                        \lim_{k^2\to 0} \, \ln\left( k^2 \right) \, \mathcal{D}(0)  \, + \,
                                  \int_{0}^{\infty} \d x \, \left[ \, x \, \ln\left( x \right) \,-\,  x\, \right]
                                            \, \mathcal{D}''(x) \, \Biggr\}\,.
\end{equation}
Clearly, we encounter a small-momentum singularity proportional to $- \mathcal{D}(0) \ln\left( k^2 \right)$, which can only be avoided upon demanding that $\mathcal{D}(0)=0$. This is the first indication that it might be difficult to have a decoupling/massive gluon, $\mathcal{D}(0)>0$, in 2 space-time dimensions. In specific cases, the infrared sickness of $d=2$ Landau-gauge theory with a massive gluon was already noticed in \cite{Dudal:2008xd,Tissier:2011ey}. The preference for the scaling solution in $d=2$ was very recently also proclaimed from a somewhat different viewpoint in \cite{Weber:2011nw}.

\subsection{Infrared singularity in and bounds on $\sigma(k^2)$ using the ghost Dyson-Schwinger equation }
Since the analysis so far is purely based on one-loop perturbation theory for the ghost propagator, albeit with a completely general gluon propagator, we need to certify our conclusion beyond this approximation. Therefore, we shall first derive a set of bounds for the one-loop result (\ref{sigma1}), valid for general $d$. We need that (see \cite{new, Lerche:2002ep})
\begin{eqnarray}
\int \, \frac{ 1 - \cos^2(\phi_1) }{  \left[ \,
                          k^2 \,+\, q^2 \, - \, 2 \, k \, q \, \cos(\phi_1) \, \right]^{\nu} } \, \d\Omega_d
  & = & \frac{\Omega_d}{q^2} \, \frac{d-1}{d} \; _2F_1\left(\nu, \nu-d/2; 1+d/2; k^2/q^2\right)\theta(q^2-k^2)\nonumber\\
 && \hspace{-7mm}+\frac{\Omega_d}{k^2} \, \frac{d-1}{d} \; _2F_1\left(\nu, \nu-d/2; 1+d/2; q^2/k^2\right)\theta(k^2-q^2)
\label{eq:2F1-uno}
\end{eqnarray}
where the Gauss hypergeometric function $_2F_1(a,b;c;z)$ is defined for $|z|<1$ by the series
\begin{equation}
_2F_1(a,b;c;z) \, = \, \sum_{n=0}^{\infty} \, \frac{(a)_n \, (b)_n}{(c)_n} \, \frac{z^n}{n!}
               \, = \, 1 \, + \, \frac{a \, b}{c} \, z \, + \, \frac{a (a+1) b (b+1)}{c (c+1)} \, \frac{z^2}{2} \, \ldots \; ,
\label{eq:serie}
\end{equation}
with $(a)_n = \Gamma(a+n)/\Gamma(a)$ the Pochhammer symbol. The previous series is converging for
$c \neq 0, -1, -2, \ldots$, and in addition also for $|z|=1$ if $\mathcal{R}e(c-a-b) > 0$. In the latter case, we have for $z=1$
\begin{equation}
_2F_1\left(a,b;c;1\right) \, = \, \frac{\Gamma(c) \, \Gamma(c-a-b)}{
                                            \Gamma(c-b) \, \Gamma(c-a)} \; .
\label{eq:gausstheorem}
\end{equation}
We also introduced the usual solid angle integral
\begin{equation}
\Omega_d \, = \, \int \, \d\Omega_d \, = \, \frac{2 \, \pi^{d/2}}{\Gamma\left(\frac{d}{2}\right)} \,.
\label{eq:omegaD}
\end{equation}
For $d=3,4$ and with $z \in [0,1]$, it can be shown that the hypergeometric function $_2F_1(1,1-d/2;1+d/2;z)$ has its maximum value, equal to 1, at $z=0$, and
its minimum value, equal to $d/(2(d-1))$, at $z=1$. For $d=2$, the function $_2F_1\left(1,1-d/2;1+d/2;z\right) \, = \, _2F_1\left(1, 0; 2; z\right)$
is identically 1 for $|z|\leq1$. As such, for general $d$, we can propose the following estimate
\begin{equation}
\frac{d}{2 \, (d - 1)} \, I_d \, \leq \, \frac{\sigma(k^2)}{g^2 N} \, \leq \, I_d \; ,
\label{eq:ineqId}
\end{equation}
with
\begin{equation}
I_d \, = \, \frac{\Omega_d}{(2\pi)^d} \, \frac{d-1}{d} \, \int_0^{\infty} \, \d q \, q^{d-1} \, \mathcal{D}(q^2) \,
                                      \left[ \, \frac{\theta(k^2 - q^2)}{k^2} \; + \, \frac{\theta(q^2 - k^2)}{q^2} \, \right] \; ,
\label{eq:intd}
\end{equation}
for the one-loop ghost self-energy
\begin{eqnarray}
\frac{\sigma(k^2)}{g^2 N} & = & \int_0^{\infty} \, \d q \, \frac{q^{d-1}}{(2\pi)^d} \, \mathcal{D}(q^2) \, \int \, \d\Omega_d \,
                          \frac{1 - \cos^2(\phi_1)}{ k^2 \,+\, q^2 \, - \, 2 \, k \, q \, \cos(\phi_1) } \label{eq:sigmaDini} \nonumber\\[2mm]
                          & = & \frac{\Omega_d}{(2\pi)^d} \, \frac{d-1}{d} \,
                                 \int_0^{\infty} \, \d q \, q^{d-1} \, \mathcal{D}(q^2) \,
                           \left[ \, \frac{\theta(k^2 - q^2)}{k^2} \, _2F_1\left(1, 1-d/2; 1+d/2; q^2/k^2\right) \right. \nonumber \\[2mm]
                  & & \left. \qquad \qquad \qquad \qquad \; + \,
                                   \frac{\theta(q^2 - k^2)}{q^2} \, _2F_1\left(1, 1-d/2; 1+d/2; k^2/q^2\right) \, \right] \;.
\label{eq:sigmaD}
\end{eqnarray}
From the bound (\ref{eq:ineqId}) we derive an equality for $d=2$, in which case there is a singularity, since, as already discussed in the previous section, $I_d$ displays one.

To go beyond perturbation theory, we take a look at the Dyson-Schwinger equation for the ghost propagator \cite{Aguilar:2004sw,Alkofer:2000wg,Atkinson:1997tu} for the $d=2$ Landau gauge
\begin{equation}
\frac{1}{\mathcal{G}(k^2)} \, = \, k^2 \, - \, g^2 \, N \,
                                  \int \frac{\d^2 q}{(2\pi)^2} \, p_{\lambda} \, \Gamma_{\lambda \nu}(k,q) \,
                                                k_{\nu} \, \mathcal{D}(q^2) \, P_{\mu \nu}(q) \, \mathcal{G}(p^2) \; ,
\label{eq:GiniDSE}
\end{equation}
where $p=k-q$ and where the gluon and the ghost propagators, $\mathcal{D}(k^2)$ and $\mathcal{G}(k^2)$, are now fully dressed propagators. The quantity $ i g f^{adc} k_{\lambda} \Gamma_{\lambda \nu}(k,q) $ corresponds to the full ghost-gluon vertex. Upon using eq.~(\ref{eq:finalG}), we can consider
\begin{equation}
\sigma(k^2) \, = \, \frac{g^2 \, N}{k^2} \,
                                  \int \frac{\d^2 q}{(2\pi)^2} \, p_{\lambda} \, \Gamma_{\lambda \nu}(k,q) \,
                                    k_{\nu} \, \mathcal{D}(q^2) \, P_{\mu \nu}(q) \, \frac{1}{p^2} \frac{1}{1 - \sigma(p^2)} \; .
\label{eq:sigmafull}
\end{equation}
For a tree-level ghost-gluon vertex $\Gamma_{\lambda \nu}(k,q) = \delta_{\lambda \nu}$ and using
the transversality of the gluon propagator we write
\begin{equation}
\sigma(k^2) \, = \, g^2 \, N \, \frac{k_{\mu} k_{\nu}}{k^2} \, \int \frac{\d^2 q}{(2\pi)^2} \,
                                    \mathcal{D}(q^2) \, P_{\mu \nu}(q) \, \frac{1}{p^2} \frac{1}{1 - \sigma(p^2)} \; ,
\label{eq:sigma2}
\end{equation}
or, using a suitable base for polar coordinates,
\begin{equation}
\frac{\sigma(k^2)}{g^2 N} \, = \, \int_0^{\infty} \frac{q \, \d q}{4 \pi^2} \, \mathcal{D}(q^2) \,
                                         \int_0^{2 \pi} \d\theta \,
                                          \frac{1 - \cos^2(\theta)}{p^2 \, \left[ 1 - \sigma(p^2) \right]} \; ,
\label{eq:sigmaDS}
\end{equation}
with $p^2 = k^2 + q^2 - 2\,k\,q\,\cos(\theta)$.

Let us first assume there is no ghost enhancement, i.e.~that $\sigma(0) < 1$. In such case, we easily derive a lower and an upper bound for the l.h.s.\ of Eq.\ (\ref{eq:sigmaDS}) as follows\footnote{Here, we assume that $\sigma(k^2)$ attains it maximum value at $k^2=0$, which was already done by Gribov \cite{LENINGRAD-77-367}. In some cases this can be shown \cite{new}, but going into those details would lead us too far now. In fact, even if $\sigma(k^2)<1$ would be maximal at $k^2=k^2_\ast>0$, the following argument would still work out by replacing $\sigma(0)$ with $\sigma(k_\ast^2)$.}:
\begin{equation}
I \, \leq \, \frac{\sigma(k^2)}{g^2 N} \, \leq \, \frac{I}{1 - \sigma(0)} \; ,
\label{eq:lowerupper}
\end{equation}
where
\begin{equation}
I \, = \, \int_0^{\infty} \frac{q \, \d q}{4 \pi^2} \, \mathcal{D}(q^2) \,
                                         \int_0^{2 \pi} \d\theta \,
                                          \frac{1 - \cos^2(\theta)}{p^2} \; .
\label{eq:I0}
\end{equation}
This integral $I$ is the one-loop ghost-self-energy contribution we already encountered before, so we know that
$I$ will display an infrared  singularity proportional to $-\lim_{k^2\to0}\ln(k^2)$ in $d=2$ unless $\mathcal{D}(0) = 0$. A fortiori, the $\sigma(k^2)$ of eq.~(\ref{eq:sigmaDS}) has a logarithmic infrared singularity, unless $\mathcal{D}(0)=0$.

A little more care is needed when $\sigma(0)=1$. Let us then write $1-\sigma(k^2)\approx ck^{2e}$ for small $k^2$ and analyze
\begin{eqnarray}
\frac{\sigma(k^2)}{g^2 N} & = & \int_0^{\infty} \frac{q \, \d q}{4 \pi^2} \, \mathcal{D}(q^2) \,
                                         \int_0^{2 \pi} \d\theta \,
                                          \frac{1 - \cos^2(\theta)}{c \, p^{2+2e}}  \nonumber\\&& + \, \int_0^{\infty} \frac{q \, \d q}{4 \pi^2} \, \mathcal{D}(q^2) \,
                                         \int_0^{2 \pi} \d\theta \, \frac{1 - \cos^2(\theta)}{p^2} \, \left[ \,
                                          \frac{1}{1 - \sigma(p^2)} \, - \, \frac{1}{c \, p^{2e}} \, \right] \; .
\end{eqnarray}
The quantity in square brackets in the last integral is finite at $p=0$ if the behavior of
$\sigma(k^2)$ is given by $1 - c k^{2e} + {\cal O}(k^b)$ with $b \geq 4e$. This quantity goes to 1
at large momenta and its absolute value is bounded from above by some positive constant $M$ if
$\sigma(k^2) \in [0,1]$. As such, we majorate as follows
\begin{equation}
\frac{\sigma(k^2)}{g^2 N} \, \leq \, \int_0^{\infty} \frac{q \, \d q}{4 \pi^2} \, \mathcal{D}(q^2) \,
                                         \int_0^{2 \pi} \d\theta \,
                                          \frac{1 - \cos^2(\theta)}{c p^{2+2e}} \, + \, M \, I \; ,
\label{eq:sigmaupper}
\end{equation}
with $I$ the same integral as before. For the other integral we can use the general result (\ref{eq:2F1-uno}) for  $d=2$ and $\nu = 1+e$, obtaining
\begin{eqnarray}
&&\int_0^{\infty} \frac{q \, \d q}{4 \pi^2} \, \mathcal{D}(q^2) \,
                                         \int_0^{2 \pi} \d\theta \,
                                          \frac{1 - \cos^2(\theta)}{c p^{2+2e}}\nonumber\\&=&
\int_0^{k} \frac{q \, \d q}{4 \pi c} \, \frac{\mathcal{D}(q^2)}{k^2} \, _2F_1\left(1+e, e; 2; q^2/k^2\right) \, + \,
\int_k^{\infty} \frac{q \, \d q}{4 \pi c} \, \frac{\mathcal{D}(q^2)}{q^2} \, _2F_1\left(1+e, e; 2; k^2/q^2\right) \; .
\label{eq:int+int}
\end{eqnarray}
From the series representation (\ref{eq:serie}) it follows that the derivative $_2F_1(a,b;c;z)$ can be brought in the following form:
\begin{equation}
\frac{\partial}{\partial z} \, _2F_1(a,b;c;z) \, = \, \frac{a \, b}{c} \, _2F_1(a+1,b+1;c+1;z) \; .
\end{equation}
In particular,
\begin{equation}
\frac{\partial}{\partial z} \, _2F_1(1+e, e; 2;z) \, = \, \frac{(1+e) \, e}{2} \,
                               _2F_1(2+e, 1+e; 3;z) \; ,
\end{equation}
i.e.\ it is positive, implying that this hypergeometric function attains its largest value at $z=1$. As $\Re(c-a-b) = 1-2e$, we have that for $e < 1/2$, the hypergeometric function is
finite at $z=1$, with a value given by $ \Gamma(1-2e)/\left[\Gamma(2-e) \, \Gamma(1-e) \right]$. Thus, for $e < 1/2$ a
singularity can only appear for $k^2 \to 0$.  Using the trapezoidal rule,
\begin{equation}
\int_a^b \,\d x f(x) \, = \, \frac{b-a}{2} \, \left[ \, f(b) + f(a) \, \right] \, + \, {\cal O}(b-a)^3 \; ,
\end{equation}
we subsequently find in the limit $k^2 \to 0$
\begin{eqnarray}\label{62}
\lim_{k^2 \to 0} \left[ \,
\frac{k^2}{8 \pi c} \, \frac{\mathcal{D}(k^2) \, _2F_1\left(1+e, e; 2; 1\right)}{k^2} \, + \,
\int_k^{\infty} \frac{\d q}{4 \pi \, c} \, \frac{\mathcal{D}(q^2)}{q} \, \right] \; .
\end{eqnarray}
As before, we note the presence of an infrared singularity, due to the last term, whenever $\mathcal{D}(0)>0$. Both terms on the r.h.s.~of the inequality
(\ref{eq:sigmaupper}) develop a $-\ln(0)$ singularity proportional to $\mathcal{D}(0)$. Since the lower bound in eq.\ (\ref{eq:lowerupper}) always applies to eq.\ (\ref{eq:sigmaDS}) and $I$ displays the same singularity as obtained for the upper bound, we are forced to conclude once more that $\sigma(k^2)$ can be finite at $k^2=0$ only if $\mathcal{D}(0)=0$.

To arrive at the general conclusion that in $d=2$ we must have $\mathcal{D}(0)=0$ in order to avoid an infrared singularity in the ghost self energy, we had to assume that $\sigma(k^2) = 1 - c k^{2e} + {\cal O}(k^b)$ with $1 > 2e$ and $b \geq 4e$ in the case ghost enhancement occurs. The $d=2$ lattice data \cite{Maas:2007uv}
suggest for the ghost propagator an infrared exponent $e \approx 0.15$, in reasonable agreement with the
scaling solution of \cite{Lerche:2002ep,Zwanziger:2001kw,Huber:2007kc} that predicted $e=1/5$.
Both lattice and analytical estimates are such that the condition $1 > 2e$ is satisfied. The demand on the subleading  exponent, $b \geq 4e$, is something we must assume, since there are neither lattice nor analytical estimates for this guy.

The attentive reader shall have noticed that so far, we relied on the use of a bare (tree level) ghost-gluon vertex. Such approximation is usually made in Dyson-Schwinger studies. Though, to obtain a qualitatively precise comparison between the lattice and analytically-obtained Dyson-Schwinger ghost propagator, one should go beyond a tree-level vertex \cite{RodriguezQuintero:2011au}. Fortunately, one recognizes that, by taking in Eq.\ (\ref{eq:sigmafull}) the full vertex
$\Gamma_{\lambda \nu}(k,q)$, instead of the tree-level one $\delta_{\lambda \nu}$, the above derived results still
apply, as long as the vertex itself remains infrared finite, as seems to be confirmed by lattice data \cite{Cucchieri:2004sq,Cucchieri:2006tf}. This illustrates the power of the presented bound analysis.

\section{What about $d=3$ or $d=4$?}
Having done all the efforts to analyze the $d=2$ ghost propagator from its Dyson-Schwinger equation, one could wonder if not a similar result could be derived in $d=3$ or $d=4$ and, if so, if we can draw any conclusions as in $d=2$? For the sake of presentation, let us stick to the $d=4$ case in this proceeding; the $d=3$ case is completely analogous. The following analysis is again originating from \cite{new}.

The Dyson-Schwinger equation for $\sigma(k^2)$ is given by \cite{Atkinson:1997tu}
\begin{equation}\label{eq:DSE4D}
\sigma(k^2)  \, = \, 1 \, - \, \widetilde{\mathcal{Z}}_3 \, + \, \widetilde{\mathcal{Z}}_1
                           g^2N\int_0^{\infty} \, \d q \, \frac{q^3}{(2\pi)^4} \, \mathcal{D}(q^2) \, \int \, \d\Omega_d \,
                                     \frac{1 - \cos^2(\phi_1)}{ p^2 \, \left[ \, 1 \, - \, \sigma(p^2) \, \right]} \; ,
\end{equation}
where $\widetilde{\mathcal{Z}}_3$ and $\widetilde{\mathcal{Z}}_1$ are the renormalization constants for the
ghost propagator and the ghost-gluon vertex respectively. We have again opted for a tree-level
ghost-gluon vertex here, but as for the $d=2$ case, if the vertex is infrared finite all results generalize to a nontrivial vertex.

Thanks to the the non-renormalization of the
ghost-gluon vertex in Landau gauge \cite{Taylor:1971ff}, we may set  $\widetilde{\mathcal{Z}}_1=1$. We shall build the argument on
\begin{equation}\label{4Da1}
\frac{\sigma_d(k^2)}{g^2N}  \, \equiv \, \int_0^{\infty} \, \d q \, \frac{q^{d-1}}{(2\pi)^d} \, \mathcal{D}(q^2) \, \int \, \d\Omega_{d} \,
                                     \frac{1 - \cos^2(\phi_1)}{ p^2 \, \left[ \, 1 \, - \, \sigma_d(p^2) \, \right]} \; ,
\end{equation}
for $d$ general, in the spirit of dimensional regularization. We will show that no infrared singularities occur for $2<d\leq 4$, consequently the ultraviolet divergences when $d\to4$ are treated by the $\mathcal{Z}$-factor(s). We again derive bounds, but this time for $\sigma_d(k^2)$. We have
\begin{equation}
\frac{\sigma_d(k^2)}{g^2 N} \, \leq \,  \frac{I(0)}{1 \, - \, \sigma_d(0)} \; ,
\label{eq:sigmaupper4d0}
\end{equation}
if $\sigma_d(k^2) \leq \sigma_d(0) < 1$, and
\begin{equation}
\frac{\sigma_d(k^2)}{g^2 N} \, \leq \, I(e) \, + \, M \, I(0) \; ,
\label{eq:sigmaupper4d}
\end{equation}
if $\sigma_d(k^2) \leq \sigma_d(0) = 1$ with $ \sigma_d(k^2) \approx 1 - c k^{2e}$ for small $k^2$. We introduced
\begin{equation}
I(e)  \, = \, \int_0^{\infty} \, \d q \, \frac{q^{d-1}}{(2\pi)^d c} \, \mathcal{D}(q^2) \, \int \,\d\Omega_d \,
                        \frac{1 - \cos^2(\phi_1)}{ (p^2)^{1+e}}\,.
\end{equation}
The basic inequalities (\ref{eq:ineqId}) can again be used to write down that
\begin{equation}
\frac{\sigma_d(k^2)}{g^2 N} \, \leq \, \frac{I_d}{1 \, - \, \sigma_d(0)} \; ,
\end{equation}
if $\sigma_d(k^2) \leq \sigma_d(0) < 1$, and
\begin{equation}
\frac{\sigma_d(k^2)}{g^2 N} \, \leq \, I(e) \, + \, M \, I_d \; ,
\end{equation}
if $\sigma_d(k^2) \leq \sigma_d(0) = 1$. The integral $I_d$ has been defined already in eq.~(\ref{eq:intd}), making clear that
$I_d$ for $2<d\leq 4$  is finite, even if $\mathcal{D}(0) \neq 0$. We can also evaluate $I(e)$ further, yielding
\begin{eqnarray}
I(e)&=&\int_0^{k} \frac{q^{d-1} \, \d q}{(2\pi)^d c}\frac{2\pi^{d/2}}{\Gamma(d/2)} \, \frac{\mathcal{D}(q^2)}{k^2} \, _2F_1\left(1+e, 1+e-d/2; 1+d/2; q^2/k^2\right)\nonumber\\ && + \,
\int_k^{\infty} \frac{q^{d-1} \, \d q}{(2\pi)^d c} \, \frac{\mathcal{D}(q^2)}{q^2}\frac{2\pi^{d/2}}{\Gamma(d/2)} \, _2F_1\left(1+e, 1+e-d/2; 1+d/2; k^2/q^2\right) \; .
\label{eq:int+int4D}
\end{eqnarray}
If $e\leq d/2-1$, the hypergeometric function attains it maximum value, $1$, at $z=0$, while the above integrals are all well-behaved for $k^2\to0$, as we assumed $d>2$.  If, on the other hand, $e>d/2-1$, the hypergeometric functions'~maximum is at $z=1$, with convergence if $e < (d-1)/2$. We can thus state that
\begin{equation}
\frac{\sigma_d(k^2)}{g^2 N} \, \leq \, \, M' \, I_4
\end{equation}
for some $M'>0$. Hence, for $d=4$ we have a finite $\sigma_d(0)$ if $\mathcal{D}(0)$ is also
finite, but it does not have to vanish, in contrast with our $d=2$ findings.

For completeness, we should mention that when $e>d/2\,-1$, in which case the hypergeometric function $_2F_1\left(1+e, 1+e-d/2; 1+d/2; z\right)$ would not be convergent at $z=1$, we are unable to derive a sensible upper-bound\footnote{Notice that this does not imply anything for $\mathcal{D}(0)$, as we would then simply prove the rather obvious bound  $\sigma_d(k^2)\leq\infty$.}. Such values for the exponent $e$ would correspond to a ghost propagator which would be very infrared singular, in particular for $d=4$, the ghost would need to be at least as singular as $1/k^{5}$. Such a strong singularity has never been proposed to our knowledge, neither numerically nor analytically.

\section{What do our $d=2$ results imply?}
We have thus analytically shown, using as sole input the ghost Dyson-Schwinger equation and the Gribov no-pole condition, that in $d=2$, one can only find a gluon propagator that vanishes at zero momentum, $\mathcal{D}(0)=0$. If $\mathcal{D}(0)>0$, a logarithmic infrared singularity occurs in $\sigma(k^2)$, which makes it impossible to fulfill the Gribov no-pole condition. The latter serves as a diagnostic tool to check whether one did not cross the first Gribov horizon, where the Faddeev-Popov operator vanishes. The used mathematical tools are easy to follow step by step, and do only rely on some general properties of e.g.~the gluon propagator, rather than on (un)controllable approximations. Our results in a way provide the analytical ``reason'' behind the $d=2$ large-volume lattice results and their accompanying infinite-volume limit \cite{Cucchieri:2007md,Cucchieri:2007rg,Maas:2007uv,Cucchieri:2011um}. Indeed, there is no sign of a decoupling/massive scenario using lattice simulations of $d=2$ Landau-gauge theory.

Our results indicate that the finding of the decoupling/massive solution, with $\mathcal{D}(0)>0$, as reported in \cite{Zwanziger:2001kw,Huber:2007kc}, should be reconsidered. Moreover, we believe they also pose a challenge for the interpretation presented in works like \cite{Maas:2009se,Maas:2010wb,Maas:2011ba} (see also \cite{Cucchieri:2011um} for a critical assessment): using the fact that one can find an infinite number of solutions to the Dyson-Schwinger equations \cite{Fischer:2008uz,Boucaud:2011ug}, it was (and still is) speculated that choosing one of these solutions would correspond to a supplementary, nonperturbative gauge-fixing choice, on top of the underlying Landau gauge; the associated gauge parameter would be the inverse of the ghost form factor at zero momentum, or equivalently in our language, the value of $\sigma(0)$. Moreover, these different gauge-fixing conditions should be related \cite{Maas:2009se,Maas:2010wb,Maas:2011ba} to different Gribov copies. If this would really be a matter of gauge fixing, it looks rather suspicious that an a priori general framework of gauge fixing would make no sense in $d=2$, where any decoupling, in the sense of $\mathcal{D}(0)>0$, is excluded. The possibility of a $d=2$ decoupling/massive solution would nevertheless be expected in the philosophy of \cite{Maas:2009se,Maas:2010wb,Maas:2011ba} (see again \cite{Cucchieri:2011um} for a critical discussion).

\section{From gluon to ghost propagator}
Armed with the current high precision lattice data for ghost and gluon propagators, it remains the question if these can be described in a single analytical framework, not only qualitatively, but also quantitatively. An interesting approach to this problem is the philosophy of \cite{Boucaud:2011ug}: given a quantitatively fine description of the lattice gluon propagator, does the corresponding solution of the ghost Dyson-Schwinger equation produce an equally fine quantitative description of the corresponding lattice ghost propagator, this without changing any of the input parameters. It was found in \cite{Boucaud:2011ug} this is not a trivial requirement, as depending on the strength of coupling constant, a set of solutions can be found. Of course, in principle the coupling constant in a given renormalization scheme is a fixed quantity. A similar observation has been made in \cite{new}. Also the role of the ghost-gluon vertex cannot be underestimated for quantitative results \cite{RodriguezQuintero:2011au}.

We follow a slightly different route, based on the papers \cite{new,new2}. In \cite{Cucchieri:2011ig,Cucchieri:2012gb}, high quality fits were presented to the lattice gluon propagator in $d=2,3,4$. In the particular case of $d=3,4$, the used fitting functions are in a 1-1 correspondence with the tree-level propagators of the Refined Gribov-Zwanziger (RGZ) formalism \cite{Dudal:2008sp,Dudal:2011gd}, thereby fixing all the input (massive) parameters of the RGZ approach using lattice input. These masses correspond to condensates, which existence is favoured based on dynamical reasons \cite{Dudal:2011gd}. Said otherwise, they correspond to another vacuum state with lower energy. The natural question arising then is whether perturbation theory around this nonperturbative vacuum is sufficient to capture all relevant dynamics for e.g.~the ghost propagator. Work in this direction is currently being undertaken in \cite{new2}. Preliminary results do point towards the need of including gluon-ghost vertex corrections to match the one-loop ghost propagator with its lattice counterpart over the full momentum range. Nevertheless the correct leading infrared behaviour is already recovered with one loop perturbation theory. It is also instructive to note here that the coupling constant is not a free parameter in this setting, as it ought to be related to the renormalization scale using its (one-loop) renormalization group equations in e.g.~the MOM scheme \cite{new2}.

\section*{Acknowledgments}
D.~Dudal and N.~Vandersickel are supported by the Research-Foundation
Flanders (FWO). A.~Cucchieri and T.~Mendes thank CNPq and FAPESP for
partial support. D.~Dudal thanks the organizers for the invitation to this pleasant workshop, and some of its participants for the enlightening discussions.


\begin{thebibliography}{99}
\bibitem{LENINGRAD-77-367}
  V.~N.~Gribov,
  %``Quantization of Nonabelian Gauge Theories,''
  Nucl.\ Phys.\ B\ {\bf 139}, 1 (1978).


\bibitem{135165}
I.~M.~Singer,
  %``Some Remarks on the Gribov Ambiguity,''
  Commun.\ Math.\ Phys.\ \ {\bf 60}, 7 (1978).


\bibitem{new}
A.~Cucchieri, D.~Dudal, N.~Vandersickel, arXiv:1202.1912 [hep-th].

\bibitem{Cucchieri:2007md}
  A.~Cucchieri, T.~Mendes,
  %``What's up with IR gluon and ghost propagators in Landau gauge? A puzzling answer from huge lattices,''
  PoS {\bf LAT2007},  297 (2007).

\bibitem{Sternbeck:2007ug}
  A.~Sternbeck, L.~von Smekal, D.~B.~Leinweber, A.~G.~Williams,
  %``Comparing SU(2) to SU(3) gluodynamics on large lattices,''
  PoS {\bf LAT2007}, 340 (2007).

\bibitem{Cucchieri:2007rg}
  A.~Cucchieri, T.~Mendes,
  %``Constraints on the IR behavior of the gluon propagator in Yang-Mills theories,''
  Phys.\ Rev.\ Lett.\  {\bf 100},  241601 (2008).

\bibitem{Bogolubsky:2009dc}
  I.~L.~Bogolubsky, E.~M.~Ilgenfritz, M.~Muller-Preussker, A.~Sternbeck,
  %``Lattice gluodynamics computation of Landau gauge Green's functions in the deep infrared,''
  Phys.\ Lett.\ B {\bf 676}, 69 (2009).

\bibitem{Cucchieri:2008fc}
  A.~Cucchieri, T.~Mendes,
  %``Constraints on the IR behavior of the ghost propagator in Yang-Mills theories,''
  Phys.\ Rev.\ D {\bf 78}, 094503 (2008).

\bibitem{Dudal:2010tf}
  D.~Dudal, O.~Oliveira, N.~Vandersickel,
  %``Indirect lattice evidence for the Refined Gribov-Zwanziger formalism and the gluon condensate $\braket{A^2}$ in the Landau gauge,''
  Phys.\ Rev.\ D {\bf 81}, 074505 (2010).

\bibitem{Cucchieri:2011ig}
  A.~Cucchieri, D.~Dudal, T.~Mendes, N.~Vandersickel,
  %``Modeling the Gluon Propagator in Landau Gauge: Lattice Estimates of Pole Masses and Dimension-Two Condensates,''
  arXiv:1111.2327 [hep-lat].


\bibitem{Aguilar:2004sw}
A.~C.~Aguilar, A.~A.~Natale,
  %``A Dynamical gluon mass solution in a coupled system of the Schwinger-Dyson equations,''
  JHEP {\bf 0408}, 057 (2004).

\bibitem{Dudal:2008sp}
  D.~Dudal, J.~A.~Gracey, S.~P.~Sorella, N.~Vandersickel, H.~Verschelde,
  %``A Refinement of the Gribov-Zwanziger approach in the Landau gauge: Infrared propagators in harmony with the lattice results,''
  Phys.\ Rev.\ {\bf D78}, 065047 (2008).

\bibitem{Dudal:2011gd}
  D.~Dudal, S.~P.~Sorella, N.~Vandersickel,
  %``The dynamical origin of the refinement of the Gribov-Zwanziger theory,''
  Phys.\ Rev.\ {\bf D84}, 065039 (2011).

%\cite{Boucaud:2008ji}
\bibitem{Boucaud:2008ji}
P.~Boucaud, J-P.~Leroy, A.~L.~Yaouanc, J.~Micheli, O.~Pene, J.~Rodriguez-Quintero,
  %``IR finiteness of the ghost dressing function from numerical resolution of the ghost SD equation,''
  JHEP {\bf 0806}, 012 (2008).


\bibitem{Aguilar:2008xm}
  A.~C.~Aguilar, D.~Binosi, J.~Papavassiliou,
  %``Gluon and ghost propagators in the Landau gauge: Deriving lattice results from Schwinger-Dyson equations,''
  Phys.\ Rev.\  {\bf D78},  025010 (2008).

\bibitem{Fischer:2008uz}
  C.~S.~Fischer, A.~Maas, J.~M.~Pawlowski,
  %``On the infrared behavior of Landau gauge Yang-Mills theory,''
  Annals Phys.\ {\bf 324}, 2408 (2009).


\bibitem{Alkofer:2000wg}
  R.~Alkofer, L.~von Smekal,
  %``The Infrared behavior of QCD Green's functions: Confinement dynamical symmetry breaking, and hadrons as relativistic bound states,''
  Phys.\ Rept.\  {\bf 353}, 281 (2001).

\bibitem{Maas:2007uv}
  A.~Maas,
  %``Two and three-point Green's functions in two-dimensional Landau-gauge Yang-Mills theory,''
  Phys.\ Rev.\  {\bf D75},  116004 (2007).

\bibitem{Cucchieri:2011um}
  A.~Cucchieri, T.~Mendes,
  %``The Saga of Landau-Gauge Propagators: Gathering New Ammo,''
  AIP Conf.\ Proc.\  {\bf 1343}, 185 (2011).

\bibitem{Dudal:2008xd}
  D.~Dudal, S.~P.~Sorella, N.~Vandersickel, H.~Verschelde,
  %``The Effects of Gribov copies in 2D gauge theories,''
  Phys.\ Lett.\ {\bf B680}, 377 (2009).

\bibitem{Tissier:2011ey}
  M.~Tissier, N.~Wschebor,
  %``An Infrared Safe perturbative approach to Yang-Mills correlators,''
  Phys.\ Rev.\  {\bf D84},   045018 (2011).

\bibitem{Weber:2011nw}
  A.~Weber,
  %``Epsilon expansion for infrared Yang-Mills theory in Landau gauge,''
  arXiv:1112.1157 [hep-th].

\bibitem{Lerche:2002ep}
  C.~Lerche, L.~von Smekal,
  %``On the infrared exponent for gluon and ghost propagation in Landau gauge QCD,''
  Phys.\ Rev.\ D {\bf 65}, 125006 (2002).

\bibitem{Atkinson:1997tu}
  D.~Atkinson, J.~C.~R.~Bloch,
  %``Running coupling in nonperturbative QCD. 1. Bare vertices and y-max approximation,''
  Phys.\ Rev.\ {\bf D58}, 094036 (1998).

\bibitem{Zwanziger:2001kw}
  D.~Zwanziger,
  %``Nonperturbative Landau gauge and infrared critical exponents in QCD,''
  Phys.\ Rev.\ {\bf D65}, 094039 (2002). % [hep-th/0109224].

\bibitem{Huber:2007kc}
  M.~Q.~Huber, R.~Alkofer, C.~S.~Fischer, K.~Schwenzer,
  %``The Infrared behavior of Landau gauge Yang-Mills theory in d=2, d=3 and d=4 dimensions,''
  Phys.\ Lett.\  {\bf B659 }, 434 (2008).

\bibitem{RodriguezQuintero:2011au}
  J.~Rodriguez-Quintero,
  %``The dimension-two gluon condensate, the ghost-gluon vertex and the Taylor theorem,''
  arXiv:1112.4749 [hep-ph].
  %%CITATION = ARXIV:1112.4749;%%


\bibitem{Cucchieri:2004sq}
  A.~Cucchieri, T.~Mendes, A.~Mihara,
  %``Numerical study of the ghost-gluon vertex in Landau gauge,''
  JHEP {\bf 0412}, 012 (2004). % [hep-lat/0408034].

\bibitem{Cucchieri:2006tf}
  A.~Cucchieri, A.~Maas, T.~Mendes,
  %``Exploratory study of three-point Green's functions in Landau-gauge Yang-Mills theory,''
  Phys.\ Rev.\ {\bf D74}, 014503 (2006). % [hep-lat/0605011].

\bibitem{Taylor:1971ff}
  J.~C.~Taylor,
  %``Ward Identities and Charge Renormalization of the Yang-Mills Field,''
  Nucl.\ Phys.\ {\bf B33}, 436 (1971).

\bibitem{Maas:2009se}
  A.~Maas,
  %``Constructing non-perturbative gauges using correlation functions,''
  Phys.\ Lett.\  {\bf B689}, 107 (2010).

\bibitem{Maas:2010wb}
  A.~Maas,
  %``On gauge fixing,''
  PoS {\bf LATTICE2010},  279 (2010).

\bibitem{Maas:2011ba}
  A.~Maas,
  %``On the structure of the residual gauge orbit,''
  arXiv:1111.5457 [hep-th].

\bibitem{Boucaud:2011ug}
  P.~.Boucaud, J.~P.~Leroy, A.~L.~Yaouanc, J.~Micheli, O.~Pene, J.~Rodriguez-Quintero,
  %``The Infrared Behaviour of the Pure Yang-Mills Green Functions,''
  arXiv:1109.1936 [hep-ph].

\bibitem{new2}
A.~Cucchieri, D.~Dudal, T.~Mendes, N.~Vandersickel, to appear (2012).

\bibitem{Cucchieri:2012gb}
A.~Cucchieri, D.~Dudal, T.~Mendes, N.~Vandersickel, arXiv:1202.0639 [hep-lat].

\end{thebibliography}
\end{document}